\documentclass[prd,aps,nofootinbib,superscriptaddress]{revtex4-2}
\usepackage{epsfig}
\usepackage{amsmath, slashed}
\usepackage{booktabs}

\usepackage{array}
\usepackage{appendix}
\usepackage{braket}
\usepackage{amssymb}
\usepackage{dutchcal}
\usepackage{ifsym}
\usepackage{comment}
\usepackage{tablefootnote}
\usepackage{longtable} 
\usepackage{mathtools}
\usepackage{amssymb}

\usepackage[table,xcdraw]{xcolor}

\usepackage{hhline}
\usepackage[most]{tcolorbox}
\usepackage{float}

\usepackage[normalem]{ulem}

\usepackage{bm}
\usepackage{amsfonts}
\definecolor{myGreen}{rgb}{0.2,0.72,0.2}
\definecolor{tableGray}{rgb}{0.9,0.9,0.9}
\definecolor{tableblue}{rgb}{0.9,0.9,1.0}

\renewcommand{\[}{\begin{equation}}
\renewcommand{\]}{\end{equation}}

\definecolor{pine}{rgb}{0.0, 0.5, 0.0}

\allowdisplaybreaks

\newenvironment{Presented}{\begin{quotation} \begin{center}
             PRESENTED AT\end{center}\bigskip
      \begin{center}\begin{large}}{\end{large}\end{center} \end{quotation}}

\makeatletter

    \def\CT@@do@color{%
      \global\let\CT@do@color\relax
            \@tempdima\wd\z@
            \advance\@tempdima\@tempdimb
            \advance\@tempdima\@tempdimc
    \advance\@tempdimb\tabcolsep
    \advance\@tempdimc\tabcolsep
    \advance\@tempdima2\tabcolsep
            \kern-\@tempdimb
            \leaders\vrule
                    \hskip\@tempdima\@plus  1fill
            \kern-\@tempdimc
            \hskip-\wd\z@ \@plus -1fill }
    \makeatother
\begin{document}

\title{A model for pion
collinear parton distribution function and form factor}

  \author{Simone Venturini}
\affiliation{Dipartimento di Fisica, Universit\`a degli Studi di Pavia, I-27100 Pavia, Italy}
\affiliation{Istituto Nazionale di Fisica Nucleare, Sezione di Pavia, I-27100 Pavia, Italy}

  \author{Barbara Pasquini}
\affiliation{Dipartimento di Fisica, Universit\`a degli Studi di Pavia, I-27100 Pavia, Italy}
\affiliation{Istituto Nazionale di Fisica Nucleare, Sezione di Pavia, I-27100 Pavia, Italy}

\author{Simone Rodini}
\affiliation{CPHT, CNRS, Ecole Polytechnique, Institut Polytechnique de Paris, Route de Saclay, 91128 Palaiseau, France}


\date{\today}

\collaboration{
MAP Collaboration}\thanks{The MAP acronym stands for ``Multi-dimensional Analyses of Partonic distributions''. It refers to a collaboration aimed at studying the three-dimensional structure of hadrons. }

\begin{abstract}
We developed a model for the pion light-front wave function (LFWF) that incorporates valence, sea and gluon degrees of freedom. Using the LFWF overlap representation, we  derived parametrizations for the pion parton distribution functions and the electromagnetic form factor. These parametrizations depend on two distinct sets of parameters, enabling separate fits of the longitudinal- and transverse-momentum dependences of the LFWF. The pion PDFs are extracted
 from available Drell-Yan and photon-production data  using
the xFitter framework and are found well compatible with existing extractions.
Furthermore, the fit of the electromagnetic form factor of the pion to all the available experimntal data works  quite successfully.

\end{abstract}

\maketitle
\begin{Presented}
DIS2023: XXX International Workshop on Deep-Inelastic Scattering and
Related Subjects, \\
Michigan State University, USA, 27-31 March 2023 \\
     \includegraphics[width=9cm]{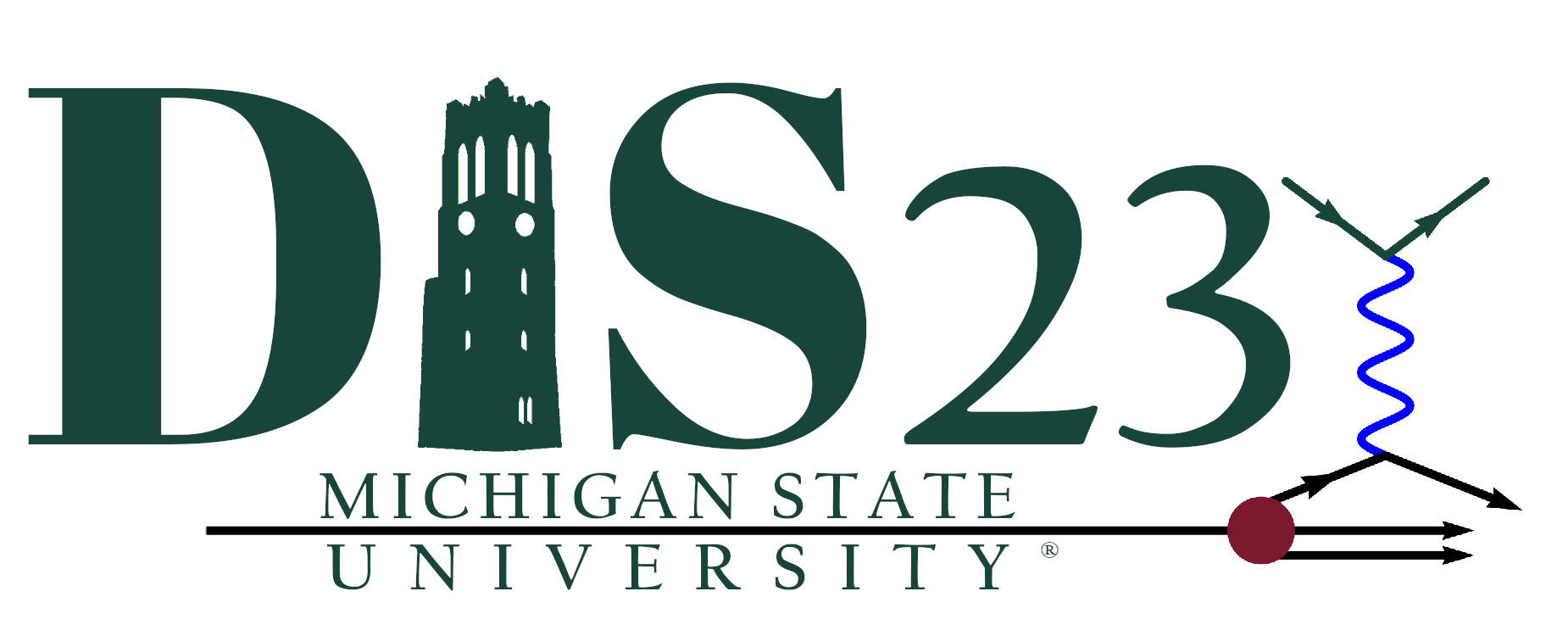}
\end{Presented}

\section{Light~-~front wave functions of the pion}
\label{Sec_Light_Front}
Light-front quantization is an effective approach in  high-energy scattering, especially for describing the hadronic matrix elements  that determine  the soft contributions in  inclusive and exclusive reactions.  These matrix elements can be expressed through the overlap of light-front wave functions (LFWFs) associated with distinct parton configurations.
Introducing the light-front Fock-state expansion and truncating the series up to the the four-partons components, the pion state can be represented as
\begin{equation}
 \label{Eq_Split_Pion_State} \ket{\pi(P)} = \ket{\pi(P)}_{q \bar{q}} + \ket{\pi(P)}_{q \bar{q}g} + \ket{\pi(P)}_{q \bar{q}gg} + \sum_{\left\{ \mathcal{s} \bar{\mathcal{s}}\right\}}\ket{\pi(P)}_{q \bar{q} \left\{ \mathcal{s} \bar{\mathcal{s}}\right\}},
\end{equation}
where $q=u,d$ and  the sum in $\left\{\bar{\mathcal{s}} \mathcal{s}\right\}$ runs over the $N_f$-flavor pairs of the sea quarks ($u\bar{u},$ $d\bar{d},$ $s\bar{s}$  at the model scale).
Restricting ourselves to consider only the contribution with zero orbital angular momentum, the LFWF for each Fock-state component in Eq.~\eqref{Eq_Split_Pion_State} can be written in a model independent way in terms of light-front wave-amplitudes (LFWAs) as~\cite{Ji:2003yj}
\begin{align}
    \label{Eq_qq_State_lz=0}
    \ket{\pi(P)}_{q \bar{q}}^{l_z = 0} &=  \int \text{d[1]}\text{d[2]}  \frac{\delta_{c_{1} c_{2}}}{\sqrt{3}}\psi^{(1)}_{q \bar{q}}(1,2) \left[q^{\dagger}_{c_{1} \uparrow}(1) \bar{q}^{\dagger}_{c_{2} \downarrow}(2) - q^{\dagger}_{c_{1} \downarrow}(1) \bar{q}^{\dagger}_{c_{2} \uparrow}(2) \right] \ket{0},
\\
    \label{Eq_qqg_State}
    \ket{\pi(P)}_{q \bar{q} g}^{l_z = 0} &= \int \text{d[1]}\text{d[2]}\text{d[3]}  \frac{T^{a}_{c_1 c_2}}{2}\psi^{(1)}_{q \bar{q} g}(1,2,3) \left[ \left(q \bar{q}\right)^{\dagger}_{A, 1} g_{a \downarrow}^{\dagger}\left(3\right) - \left(q \bar{q}\right)^{\dagger}_{A, -1} g_{a \uparrow}^{\dagger}\left(3\right) \right]\ket{0},\\
    \label{Eq_qqgg_State}
    \nonumber \ket{\pi(P)}_{q \bar{q} gg}^{l_z = 0} &=  \int \text{d[1]}\text{d[2]}  \text{d[3]} \text{d[4]}  \frac{\delta_{c_1 c_2}\delta^{a b}}{\sqrt{24}} \biggl\{\psi^{(1)}_{q \bar{q} gg}(1,2,3,4) \left(q \bar{q}\right)^{\dagger}_{A, 0} \left(gg\right)^{\dagger}_{S,0} \\
    & + \psi^{(2)}_{q \bar{q} gg}\left(1,2,3,4\right)\left(q \bar{q}\right)^{\dagger}_{S,0}\left(gg\right)^{\dagger}_{A,0}\biggr\} \ket{0},
\\
    \label{Eq_qqqq_State}
    \nonumber \ket{\pi(P)}_{q \bar{q} \left\{\mathcal{s}\bar{\mathcal{s}}\right\}}^{l_z = 0} &=  \int \text{d[1]}\text{d[2]}\text{d[3]}\text{d[4]}  \frac{\delta_{c_1 c_2}\delta_{c_3 c_4}}{3} \biggl\{\psi^{(1)}_{q \bar{q} \mathcal{s} \bar{\mathcal{s}}}(1,2,3,4) \left(q \bar{q}\right)^{\dagger}_{A, 0} \left(\mathcal{s} \bar{\mathcal{s}}\right)^{\dagger}_{S,0} \\
    \nonumber & + \psi^{(2)}_{q \bar{q} \mathcal{s} \bar{\mathcal{s}}}\left(1,2,3,4\right)\left(q \bar{q}\right)^{\dagger}_{S,0}\left(\mathcal{s} \bar{\mathcal{s}}\right)^{\dagger}_{A,0}\\
    &+ \psi^{(3)}_{q \bar{q} \mathcal{s} \bar{\mathcal{s}}}\left(1,2,3,4\right)\left[\left(q \bar{q}\right)^{\dagger}_{A,1}\left(\mathcal{s} \bar{\mathcal{s}}\right)^{\dagger}_{A,-1} - \left(q \bar{q}\right)^{\dagger}_{A,-1}\left(\mathcal{s} \bar{\mathcal{s}}\right)^{\dagger}_{A,1}\right] \biggr\} \ket{0}.
\end{align}
The measures in Eqs.~\eqref{Eq_qq_State_lz=0}-\eqref{Eq_qqqq_State} are defined as:
\begin{align}
\label{Eq_Measures}
\displaystyle \prod_{i=1}^N \text{d[}i\text{]} = \left[dx\right]_N \left[d^2 \boldsymbol{k}_{\perp}\right]_N,
\end{align}
\begin{align}
    \label{Eq_Measure_xN}
   \left[dx\right]_N = \prod_{i=1}^{N}\frac{ dx_i}{  \sqrt{ x_i}}  \delta\left(1 - \displaystyle\sum_{i=1}^{N} x_i\right), \qquad \qquad
    \left[d^2 \boldsymbol{k}_{\perp} \right]_{N} = \frac{1}{[2(2 \pi)^3]^{N-1}}\prod_{i=1}^{N}d^2 \boldsymbol{k}_{\perp i} \delta^{(2)}\left(\sum_{i=1}^{N}\boldsymbol{k}_{\perp i}\right).
\end{align}
 The LFWAs $\psi_{\beta}^{(i)}$ are generic functions depending on $i=\left(x_i, \mathbf{k}_{\perp i}\right)$, where $x_i$ and $\mathbf{k}_{\perp i}$ are, respectively, the fraction of pion momentum in the collinear direction carried by the $i$-th parton and the  transverse momentum of the $i$-th parton.
 Furthermore,  $q^\dagger_{c\lambda}, \bar{q}^{\dagger}_{c\lambda} $ and $g_{a \lambda}^{\dagger}$ are creation operator of quarks, antiquarks and gluons, respectively, and
 are labelled by the specific quantum numbers of the partons: the color ($c$ and $a$), the helicity ($\lambda=\uparrow$ or $\downarrow$), and, in case of the fermions, the flavor $q$. For brevity, in Eqs.~\eqref{Eq_qq_State_lz=0}-\eqref{Eq_qqqq_State} we introduced the operators $(p_1 p_2)^{\dagger}_{A/S, 0/1}$: these are anti-symmetric ($A$) or symmetric ($S$) combinations of creation operators of the partons $p_1$ and $p_2$ with total helicity $0$ or $1$. The subscript $\mathcal{s} \bar{\mathcal{s}}$ refers to the sea quarks $u\bar{u},$ $d\bar{d},$ $s\bar{s}$ and $T^a_{i j} = \frac{\lambda^a_{i j}}{2}$  represent the SU$(3)$ color matrices.\\
 The quark and gluon  PDFs of the pion are defined as
 \begin{align}
    \label{Eq_qq_correlator}
    f_1^q(x) & = \int \frac{dz^-}{2(2\pi)} e^{ik^+z^-}\langle \pi(p)| \bar{\psi}(0) \gamma^+ \psi(z) |\pi(p)\rangle\Big|_{\substack{z^+=0 \\ \mathbf z_\perp=\mathbf{0}}} ,\\
     \label{Eq_qqg_correlator}
     f_1^g(x) &= \frac{1}{xp^+}\int \frac{dz^-}{2\pi}  e^{ik^+z^-} \langle \pi(p)|G^{+\mu}(0)G_{\mu +}(z) |\pi(p)\rangle\Big|_{\substack{z^+=0 \\ \mathbf z_\perp=\mathbf{0}}} ,
 \end{align}
where $\psi$ is the quark field  and $G^{\mu\nu}$ is the gluon field strength tensor.
By inserting in Eqs.~\eqref{Eq_qq_correlator} and \eqref{Eq_qqg_correlator} the pion state of Eq.~\eqref{Eq_Split_Pion_State}, one obtains the representation of the pion PDFs in terms of overlap of LFWFs for each Fock-state component, as given in Ref.~\cite{Pasquini:2023aaf}.
Moreover, by neglecting electroweak corrections and quark masses, charge symmetry imposes $f_{1,\pi^+}^u=f_{1,\pi^+}^{\bar d}=f_{1,\pi^-}^{d}
=f_{1,\pi^-}^{\bar u}=2f_{1,\pi^0}^{u}=2f_{1,\pi^0}^{\bar u}=2f_{1,\pi^0}^{d}=2f_{1,\pi^0}^{\bar d}$.
Hereafter,
we will refer to distributions in positively charged pions.
Assuming  also a SU(3)-symmetric sea, i.e., $f_{1}^{u} = f_{1}^{\bar{d}} = f_{1}^{s} = f_{1}^{\bar{s}}$, we will consider three independent PDFs: the total valence contribution $f_{1}^{v}$, the total sea contribution $f_{1}^{S}$, given by
\begin{align}
    \nonumber f_{1}^{v}& = f_{1}^{u_{v}}-f_{1}^{d_{v}}  = \big(f_{1}^{u} - f_{1}^{\bar{u}}\big) - \big(f_{1}^{d} - f_{1}^{\bar{d}}\big)
     = 2 f_{1}^{u_{v}},\\
   f_{1}^{S}& = 2f_{1}^{u} + 2 f_{1}^{\bar{d}} + f_{1}^{s} + f_{1}^{\bar{s}}= 6 f_{1}^{u},     \label{Eq_SU3Symmetry2}
\end{align}
and the gluon contribution $f_{1}^{g}$.

As discussed in Ref.~\cite{Pasquini:2023aaf},  we can analogously obtain the LFWF overlap representation for the pion e.m. form factor, starting from the definition
\begin{align}
    F_\pi(Q^2)=\frac{1}{(p+p')^+}\langle \pi(p')| \bar{\psi}(0) \gamma^+ \psi(0) |\pi(p)\rangle
\end{align}
where $Q^2=-q^2>0$ and $q=p'-p$ is the four-momentum transfer.

\subsection{Parametrization}
    \label{Sec_Model}

We have designed  the LFWAs  in such a way that the parameters associated with longitudinal and transverse momentum dependence are treated separately during the fitting process of the pion PDFs and electromagnetic (e.m.) form factor (FF).
Specifically,
the LFWAs are written in the following general form
\begin{equation}
    \label{Eq_Factorized_LFWA}
    \psi^{(i)}_{\beta}\left(1, 2, \dots , N\right) = \phi^{(i)}_{\beta}\left(x_1, x_2, \dots , x_N\right) \Omega^{(i)}_{\beta}\left(x_1, \boldsymbol{k}_{\perp 1}, x_2, \boldsymbol{k}_{\perp 2}, \dots,  x_N, \boldsymbol{k}_{\perp N} \right),
\end{equation}
where the functions $\phi_{\beta}^{(i)}$
can be expressed
as a linear combination of pion distribution amplitudes  and $\Omega^{(i)}_{\beta}$ are
modified $x$-dependent gaussian functions.
The last ones are normalized in such a way  that the fit of the collinear PDFs does not contain any spurious dependence of the parameters in the transverse-momentum space.
In total the model has a set $\mathcal{X}$ of six parameters for the longitudinal momentum dependence, fitted to pion PDFs, and a set $\mathcal{A}$ of four parameters for the transverse momentum dependence, constrained from the fit to the pion e.m. FF.


\begin{figure}[t]
    \centering
    \includegraphics[width=0.49\textwidth]{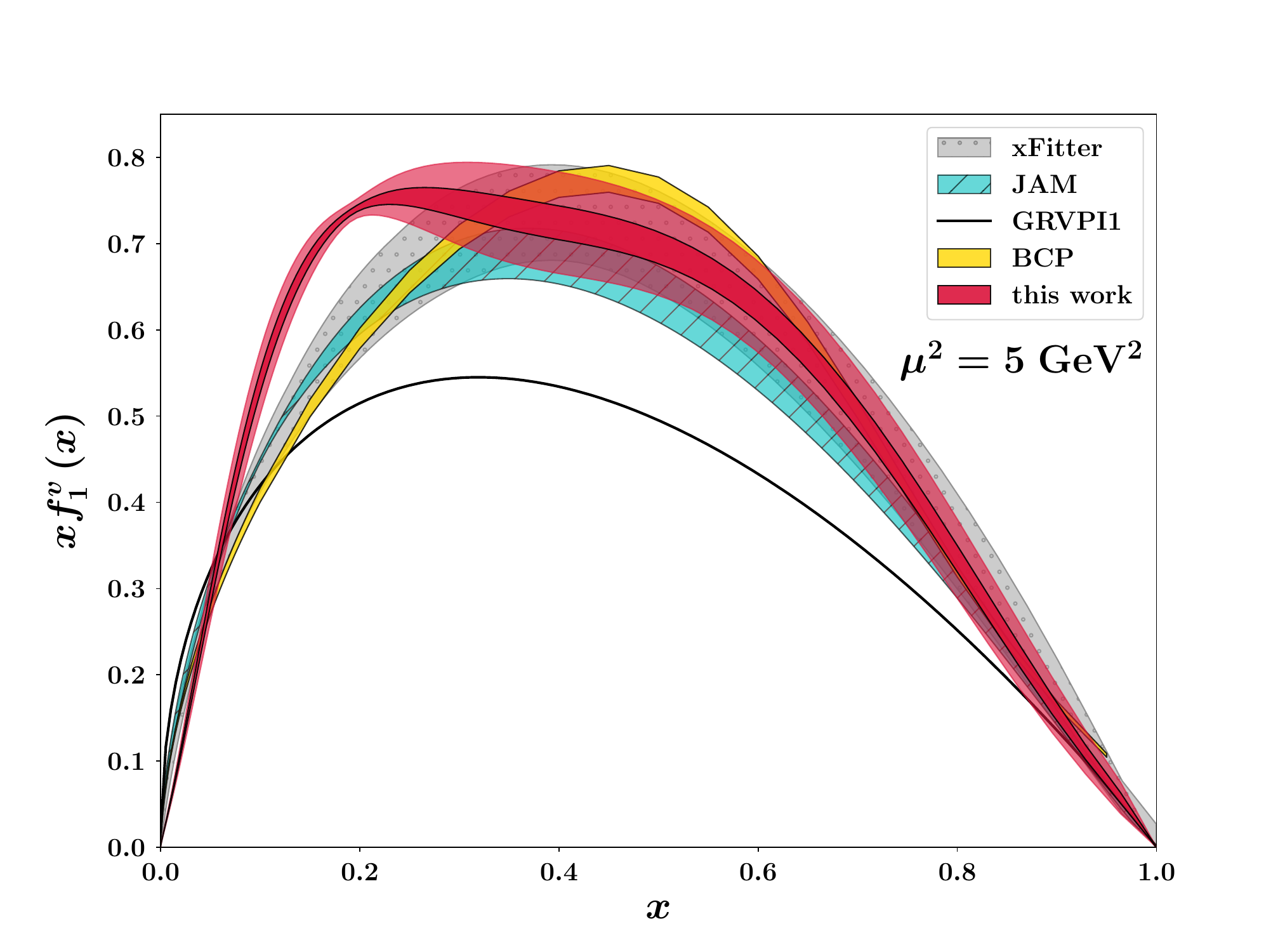}
    \\
    \includegraphics[width=0.49\textwidth]{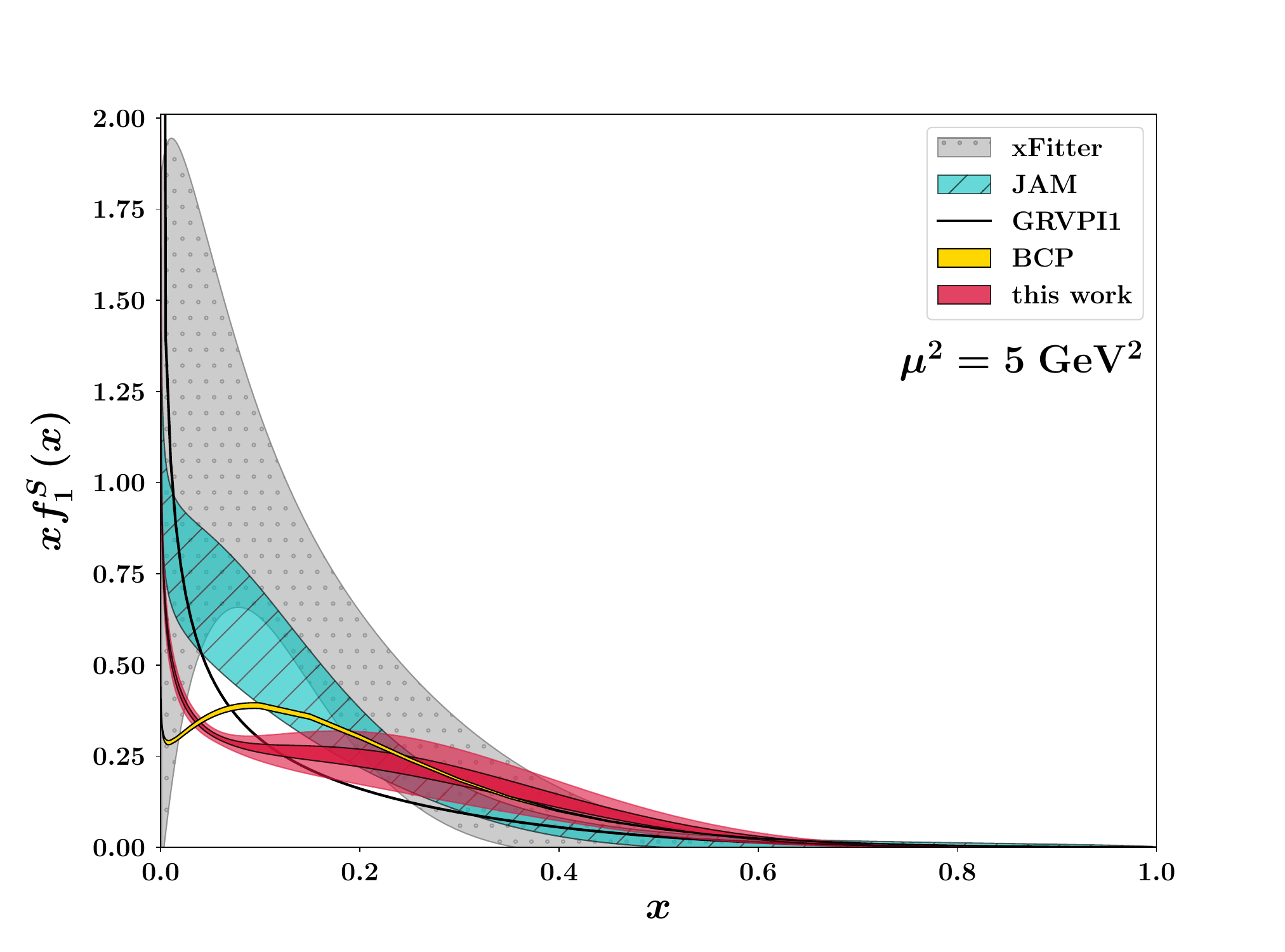}
    \includegraphics[width=0.49\textwidth]{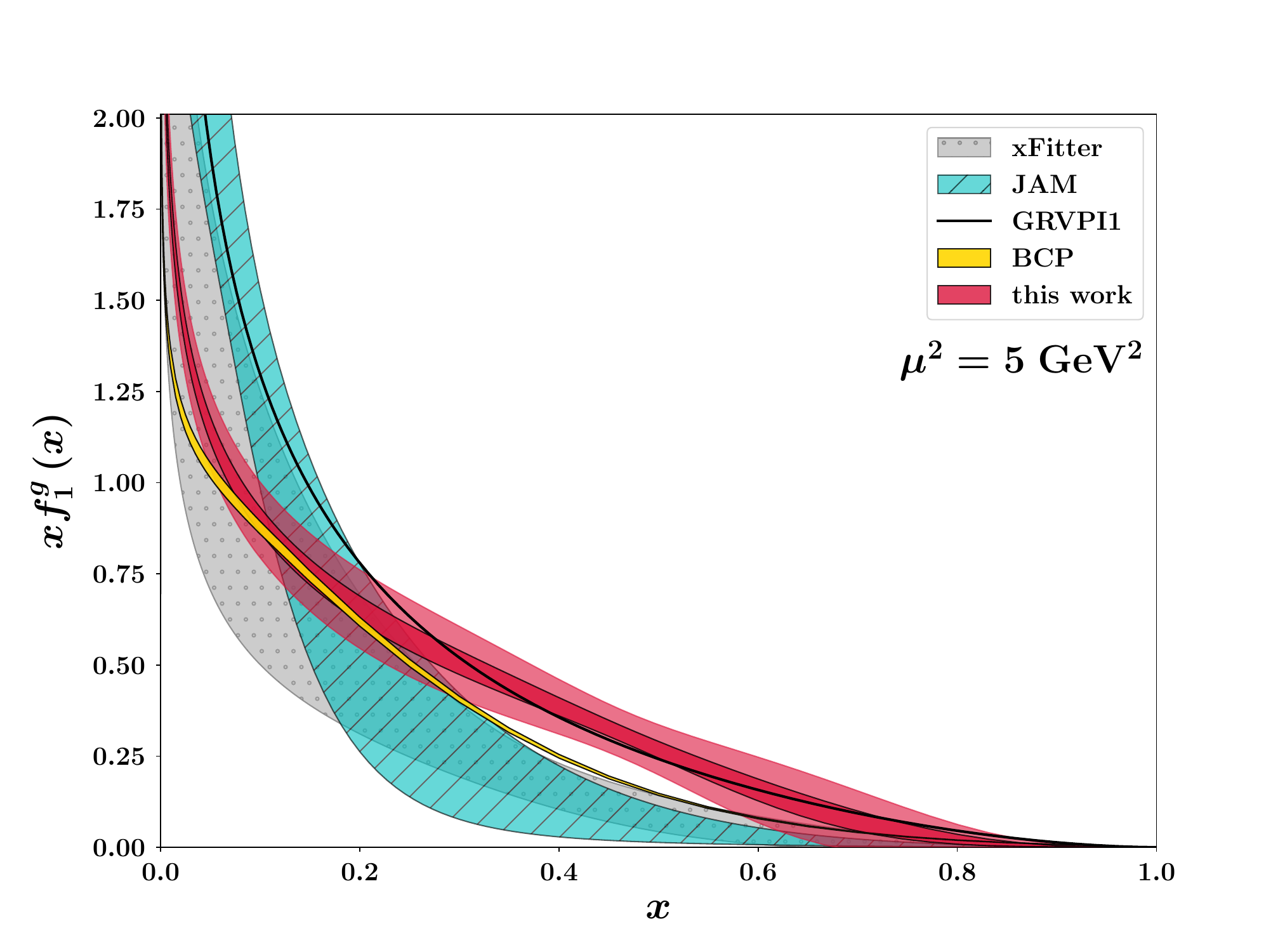}
    \caption{ $xf_{1}$ as function of $x$  for the  total valence (upper panel), total sea  (left panel in the bottom) and gluon (right panel in the bottom) contributions  at $\mu^2=5$ GeV$^2$. The light (dark) red bands show the results of this work with the 3$\sigma$  (1$\sigma$) uncertainty in comparison with the results from the JAM collaboration~\cite{Barry:2021osv} (light blue bands), the analysis of xFitter collaboration~\cite{Novikov:2020snp} (grey bands),  the BCP fit  of Ref.~\cite{Bourrely:2022mjf} (yellow bands) and  the GRVPI1 fit~\cite{Gluck:1991ey} (solid black  curves).
    \label{Fig_All_PDFs}}
\end{figure}
\section{Extraction of the pion PDF and electromagnetic form factor}
\label{Sec_Fit_and_Results}

The fit of the pion PDFs has been performed by using the open-source tool  xFitter~\cite{Alekhin:2014irh,Novikov:2020snp}. The data included in the analysis are from the
NA10~\cite{NA10:1985ibr}, E615~\cite{Conway:1989fs} and WA70~\cite{WA70:1987bai} experiments. The complete data set has been cut to exclude the kinematic region corresponding to the $J/\psi$ and $\Upsilon$ resonances.
We fixed  the initial scale to $\mu_0=0.85$ GeV and the factorization scale $\mu_F$  and renormalization scale $\mu_R$  to $\mu_F = \mu_R = 0.8 $ GeV.
The  reduced chi-squared from a single minimization is $\hat{\chi}^2/N_{d.o.f.} = 0.88$ for the number of degrees of freedom $N_{d.o.f.} = 260-6=254$.
The fit to the real data  has been repeated $1000$ times varying the  experimental points by random gaussian shifts  both for the statistic and systematic uncertainties, thus obtaining 1000 bootstrap replicas. The renormalization scale and the factorization scale have been varied replica by replica in a way such that $\mu_0 \le \mu_F \le \mu_0$ and $\mu_0 \le \mu_R \le 2\mu_0$.
\\
In Fig.~\ref{Fig_All_PDFs}  we show our results  for the pion PDFs (red bands) at the scales of $\mu^2=5$ GeV$^2$ in comparison with different analyses: the GRVPI1 solution~\cite{Gluck:1991ey} (solid black  curves);
 the xFitter results~\cite{Novikov:2020snp} (grey bands); the JAM extraction~\cite{Barry:2021osv} (light blue bands);
the results within the statistical model of Bourrely-Chang-Peng (BCP) (Bourrely~\cite{Bourrely:2022mjf}  (yellow bands).
Overall, the modern analyses give compatible results within the relative error bands. The agreement is better  for the valence and sea contributions at larger $x$ and for the gluon PDF in the small $x$ region.

\begin{figure}[ht]
    \centering
    \includegraphics[width=0.7\textwidth]{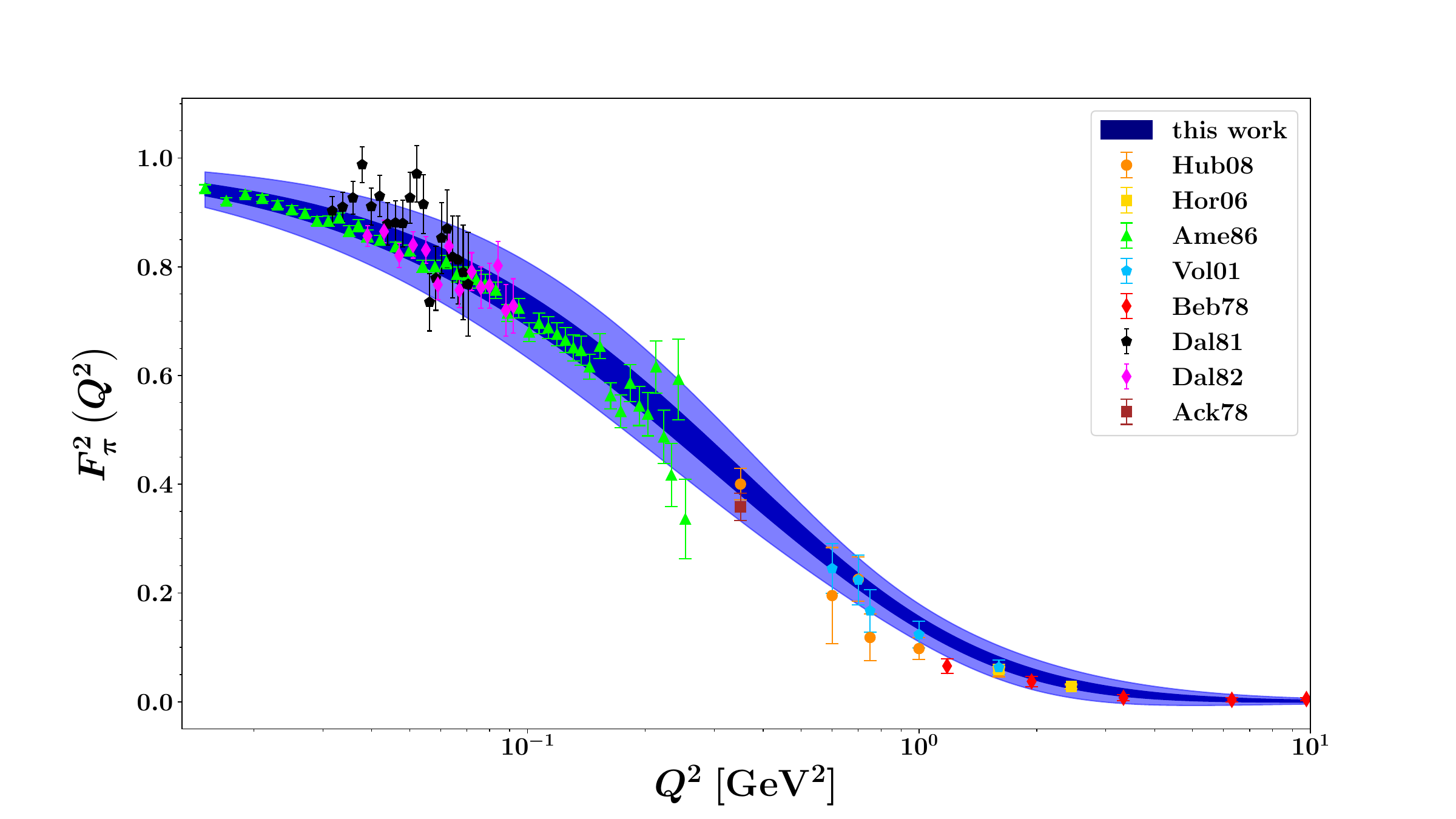}
    \caption{Fit results for the square of the pion electromagnetic form factor as function of $Q^2$.
    The dark (light) blue band shows the 68\% (99.7\%) of the replicas. The experimental data correspond to
    Hub08~\cite{JeffersonLab:2008jve},
    Hor06~\cite{JeffersonLabFpi-2:2006ysh},
    Ame86~\cite{NA7:1986vav},
    Vol01~\cite{JeffersonLabFpi:2000nlc},
    Beb78~\cite{Bebek:1977pe},
   Dal82~\cite{Dally:1981ur,Dally:1982zk},
    Ack78~\cite{Ackermann:1977rp}.
    }
    \label{Fig_FF}
    \end{figure}

For the fit of the FF, we included 100 experimental points in a $Q^2$ range from $0.015$ GeV$^2$ to $9.77$ GeV$^2$, corresponding to the experimental measurements described in the legend of
Fig.~\ref{Fig_FF}.
The FF depends both on the set
$\mathcal{X}$, fixed from the pion PDF fit,
 and the set $\mathcal{A}$, which contains four distinct elements to be determined from the FF fit.
 The bootstrap replica method is used to propagate both  the experimental uncertainties
 and the uncertainties associated to the parameters of the set $\mathcal{X}$  that are not directly free fitting variables. The  method is described in  detail in Ref.~\cite{Pasquini:2023aaf}. In the end, the best fit produces a reduced chi-squared of $1.19$.

In Fig.~\ref{Fig_FF}, we show the results for the square of the pion e.m. form factor with
the inner (dark blue) band representing the
68\% uncertainty and the external (light blue) band showing the
99.7\% uncertainty.
Agreement with the different data sets is qualitatively evident.
We stress that the two bands incorporate the error propagation of the PDF parameters, representing therefore more than just the experimental uncertainty on the FF.

\section{Conclusions}
\label{Sec_Conclusions}
In this study, we have presented an extraction of the pion PDFs and e.m. FF using a light-front model approach that incorporates contributions from valence quarks, sea quarks, and gluons. We have developed a model for the pion LFWFs, ensuring that the fit of the collinear PDFs remains free from any spurious dependence on the parameters in the transverse-momentum space. These parameters are separately fitted to the available data on the pion e.m. form factor.
The approach presented in this study serves as a proof-of-principle for achieving a unified description of hadron distribution functions, encompassing both the longitudinal and transverse momentum dynamics of partons within hadrons. The ultimate objective is to extend this framework to incorporate transverse-momentum dependent parton distributions (TMDs) and generalized parton distributions (GPDs) in a global fit. This will leverage the wealth of data expected from upcoming experiments at JLab, COMPASS++/AMBER, and future electron-ion colliders, taking the analysis a step further than previous studies that have only considered simultaneous PDFs and TMDs analysis~\cite{Barry:2023qqh}, or have focused solely on TMDs~\cite{Vladimirov:2019bfa,Cerutti:2022lmb}, and GPDs~\cite{Chavez:2021llq} separately.


\end{document}